\documentclass[12pt]{article}

\usepackage{amsmath}
\usepackage{graphicx,psfrag,epsf}
\usepackage{enumerate}
\usepackage{mathtools}
\usepackage{natbib}
\usepackage{floatrow}
\usepackage{float}
\usepackage{bm}
\usepackage[title]{appendix}
\usepackage{ulem}
\usepackage{dsfont}
\usepackage{setspace}
\usepackage[hyphens]{url}
\usepackage{hyperref}
\usepackage{caption}
\usepackage[left=1in,right=1in,top=1in,bottom=1in]{geometry}

\title{\bf An Advanced Hidden Markov Model for Hourly Rainfall Time Series}

\author{Oliver Stoner\textsuperscript{1} and Theo Economou}
\begin{document}
\maketitle
\begin{center}
Department of Mathematics, University of Exeter, UK.

\textsuperscript{1}O.R.Stoner@exeter.ac.uk
\end{center}

\begin{abstract}
For hydrological applications, such as urban flood modelling, it is often important to be able to simulate sub-daily rainfall time series from stochastic models. However, modelling rainfall at this resolution poses several challenges, including a complex temporal structure including long dry periods, seasonal variation in both the occurrence and intensity of rainfall, and extreme values. 

We illustrate how the hidden Markov framework can be adapted to construct a compelling model for sub-daily rainfall, which is capable of capturing all of these important characteristics well. These adaptations include clone states and non-stationarity in both the transition matrix and conditional models. Set in the Bayesian framework, a rich quantification of both parametric and predictive uncertainty is available, and thorough model checking is made possible through posterior predictive analyses. Results from the model are interpretable, allowing for meaningful examination of seasonal variation and medium to long term trends in rainfall occurrence and intensity. To demonstrate the effectiveness of our approach, both in terms of model fit and interpretability, we apply the model to an 8-year long time series of hourly observations.
\end{abstract}


\newpage
\section{Introduction}\label{sec:intro}
Severe flooding events, such as those that occurred in the UK in the winter of 2013-2014 and the winter of 2015-2016, pose a great risk to society. For each winter, the total economic damage caused by the flooding was estimated to be over one billion pounds (GBP) of economic damage (\cite{EA1314}, \cite{EA1516}). Hydrological flood models play an important role in mitigating this damage, for example by helping to inform the planning of new flood defences and drainage systems, as well as integration with flooding warning systems. 

Typically, hydrological models are used to test the response of the hydrological system to design storms, which are intended to represent an idealised extreme rainfall scenario. However, \cite{rainfall_uncertainty} argue that this approach is limited, in the first instance because the temporal profile of the design storm may fail to capture important characteristics of system performance. Moreover, focusing on the response of the system to a single event may be inadequate, as the risk of flooding posed by a single storm event depends strongly on the antecedent conditions of the catchment \citep{rainfall_uncertainty}. For example, the risk of flooding may depend on whether or not the catchment has already been saturated by recent rainfall. For this reason, \cite{rainfall_uncertainty} argue that hydrological models should instead use long rainfall time series generated from stochastic/probabilistic models as inputs, so that the effects of both the rainfall intensity during a storm event and antecedent conditions can be taken into account.
\begin{figure}[ht!]
\caption{Empirical quantiles of the hourly rainfall intensities (observations greater than 0mm) from the Exeter guage time series (left), empirical quantiles of dry period lengths (where rainfall does not exceed 0.2mm in any given hour), and theoretical quantiles of a Geometric fit. The points are 99\% quantiles.}\label{fig:data_quantiles}
\includegraphics[width=\linewidth]{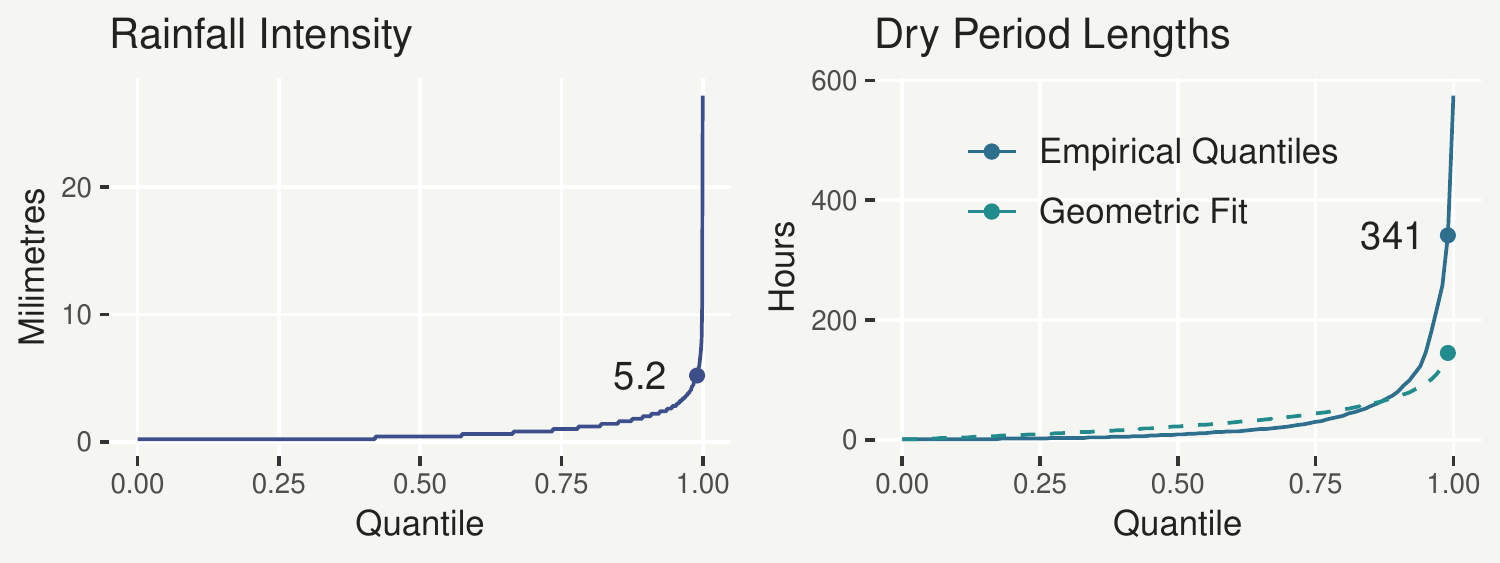}
\end{figure}

Modelling rainfall is challenging because its natural variability can dominate any seasonal structures or long-term temporal trends, more so than other meteorological variables such as temperature and wind \citep{rainfall_uncertainty}. For this reason, it is vitally important to capture well the whole distribution of rainfall values, which is in itself a non-trivial task due to the propensity of extremely high rainfall values. This is illustrated in the left plot of Figure \ref{fig:data_quantiles}, which shows empirical quantiles of non-zero rainfall gauge observations from an 8-year hourly time series in Exeter, UK \citep{data}. The heavy tail is especially clear when comparing the highest hourly values (well over 20mm) to the 99\% quantile, which is only 5.2mm. Ensuring the model is capable of reproducing these extremes is essential in any application where they are of great concern, such as surface flood modelling. An equivalent return-level plot can be found in the supplementary material.

A further modelling challenge is the complex temporal structure in the occurrence of rainfall, which notably consists of long dry periods. To illustrate this, the right plot in Figure \ref{fig:data_quantiles} shows the empirical quantiles of dry period lengths (defined as periods where the hourly rainfall value does not exceed 0.2mm). This distribution also has quite a heavy tail, including several high values which may be considered extremes. Capturing these extremes is equally important, particularly for applications where droughts are a concern, such as agricultural planning.

Finally, both rainfall intensity and occurrence can vary substantially by season. In the UK for instance, summer rainfall extremes tend to be more severe than in the winter \citep{rainfall_uncertainty}. For this reason, care should be taken to ensure model simulations accurately reflect the time of year.

In this article we propose a flexible model, based on advanced hidden Markov latent states, which is capable of capturing key features of sub-daily rainfall well:
\begin{itemize}
\item Seasonal and temporal variation in occurrence and intensity;
\item Long dry periods (which vary with time and season);
\item Extreme values (which vary with time and season).
\end{itemize}
Given that such stochastic rainfall models are often used in decision making (e.g. warnings) or as inputs to physical models (e.g. hydrological), a further requirement is that parametric uncertainty is fully quantified and propagated in the output. As such the proposed models are implemented in the Bayesian hierarchical framework. Applied to 8 years of hourly rainfall gauge data, we  demonstrate using posterior predictive model checking how our proposed model performs well overall and specifically with respect to capturing these three features.

The article is structured as follows: Section \ref{sec:background} gives an overview of existing approaches to stochastic/statistical modelling of rainfall, highlighting their strengths and limitations. In Section \ref{sec:method}, we propose a flexible model for sub-daily rainfall gauge time series, which we then apply to an 8-year long time series of hourly observations. In Section \ref{sec:check}, we present extensive model checking and an analysis of model output. Finally, the article ends with a critical discussion of our approach in Section \ref{sec:discuss}.

\section{Background}\label{sec:background}
In the previous section we outlined features of rainfall occurrence and intensity, namely long dry periods and extremely high values, which make modelling challenging. The literature on statistical modelling of rainfall is vast and comprehensive reviews are available in \cite{rainfall_uncertainty} and \cite{wilks_review}. Here we seek only to give a brief overview to establish our contribution. To this end, we focus on approaches which take into account the temporal structure of rainfall and those which model both occurrence and intensity. Most of these can be roughly separated into two classes: direct and indirect rainfall models.
 
\subsection{Direct rainfall models}
The first way of modelling rainfall time series is to characterise the occurrence and amount of rainfall at a given time as random quantities, arising directly from some probabilistic framework -- typically some form of regression.

One such approach (e.g. \cite{Yang2005}) consists of two Generalized Linear Models (GLM) \citep{GLM}, one for occurrence and one for intensity. First the data (typically daily rainfall) is transformed into a sequence of 0s (zero rainfall) and 1s (greater than zero rainfall). The model for occurrence is then a logistic regression for the probability  of rain $p_i$ on the day $i$:
\begin{equation}
\log\left(\frac{p_i}{1-p_i}\right) = \bm{x}_i'\bm{\beta}
\end{equation}
where $\bm{\beta}_i$ is a vector of coefficients corresponding to covariates $\bm{x}_i$. Then, conditional on the occurrence of rain, the mean amount of rain for wet day $j$ is modelled by a Gamma regression:
\begin{equation}
\log(\mu_j)=\bm{\xi}_j'\bm{\gamma}
\end{equation}
where $\bm{\gamma}_j$ is a vector of coefficients corresponding to covariates $\bm{\xi}_j$. The shape parameter is assumed constant.

Covariates $\bm{x}_i$ and $\bm{\xi}_j$ can include climatological variables such as the North Atlantic Oscillation (NAO), and seasonal variation can be captured through covariates such as the month. Temporal structure can be introduced in the model for occurrence by including in $\bm{x}_i$ binary variables representing the occurrence of rain in the previous $k$ days, equivalent to a $k$-order Markov model \citep{Yang2005}. Similarly, $\bm{\xi}_j$ can include the rainfall amount from previous days to induce temporal structure in rainfall intensity.

An advantage of this approach is that, set in the GLM framework, models can be implemented quickly, both in terms of the computational cost of fitting and the relatively little coding required to construct and modify them. Furthermore, these models have been extended to incorporate spatial structures, for applications to data from multiple sites \citep{Yang2005}. Further extensions involve temporally structured random effects \citep{Glasbey1997} and distributions outside the exponential family \citep{Serinaldi2012}.

However, the main downside of this approach is that the models for occurrence and intensity are separated. This may be a valid approach for coarser temporal resolutions (e.g. daily rainfall), but for sub-daily rainfall it may result in failure to capture important patterns. For example, following a period of no rainfall, the arrival of a storm might might lead to initially low rainfall intensity, which increases over a few hours before reaching a maximum and then decreasing. Being able to simulate such patterns may be essential for testing the response of a drainage system to a severe storm, for instance.

\subsection{Indirect rainfall models}\label{sec:subdaily}
An alternative approach is to instead stochastically model the larger structures which generate rainfall, such as storm events. Given the occurrence of such structures, the amount of rainfall over a time period is then deterministic, based on some simplifying assumptions (see for instance \cite{rodriguez}). Several parameters control different aspects of the model, such as storm event duration, frequency of occurrence, and intensity. Through these parameters, it is possible to tune the model to capture certain properties of the observed data, such as the length of dry periods, temporal dependence and the distribution of non-zero rainfall. 

Whilst these approaches consist of few parameters and are in some sense based on physical justification, the absence of a likelihood function on which to base inference makes implementation challenging. Thus it is often necessary to rely on alternative methods of implementation, such as by maximising objective functions (potentially missing out on parametric uncertainty) or by employing Approximate Bayesian Computation (ABC), which requires the careful selection of summary statistics \citep{ABC2}. Furthermore, it is not evident that these approaches are able to reproduce extreme rainfall values well, a feature which \cite{rainfall_uncertainty} argue is lacking in the sub-daily rainfall modelling literature. For these reasons, we pursue a fully probabilistic framework to directly model the distribution of rainfall in time. We argue that direct characterisation of the distribution constitutes a more interpretable framework, particularly in terms of model expansion and checking.

\subsection{Hidden Markov models (HMMs)}\label{sec:hmm}
A further family of direct probabilistic models that has been used extensively for rainfall data (e.g. \cite{Kim2017}, \cite{tenstates}), to capture temporal structures in both rainfall occurrence and intensity, is hidden Markov models (HMMs). In HMMs, a hidden, unobservable quantity $z_t$ varies over discrete time steps, alternating between a finite number $Z$ of values or states $z_t\in \{S_1,...S_Z\}$. Variable $z_t$ is a discrete Markov chain, whose evolution over time is probabilistic, governed by a transition matrix $P$ of probabilities. The particular state the hidden variable is in at a given time step affects (the parameters of) the conditional model for the observed quantity, which in the case of rainfall translates to the model(s) for occurrence and intensity. 

HMMs are useful for rainfall because they can capture its temporal behaviour through the Markovian structure of the latent chain, without the need to explicitly include climatological structures, such as the arrival of weather fronts, or other physical processes. However in the same vain as indirect rainfall models they offer a high degree of interpretability, in that the latent states can represent weather features such as dry periods or different stages in the arrival of storms. This is aided by the fact that occurrence and intensity are both driven by the same latent states.

Conventional HMMs for rainfall often suffer from a number of shortcomings, such as underestimation of the length of long dry periods \citep{rainfall_uncertainty}. However, their flexibility as a framework, afforded by the freedom to specify virtually any conditional model for occurrence and intensity, means that it may be possible to address these issues through a number of extensions. Based on this idea, in the subsequent section we will present our approach to modelling rainfall gauge data, which we argue is capable of capturing all the key features of rainfall data identified in Section \ref{sec:intro}.

\section{Methodology} \label{sec:method}

\subsection{Conventional HMMs for rainfall}\label{sec:conventional}
It is instructive to begin by defining a basic and generic HMM for rainfall. Capturing the whole distribution (tails and bulk) of rainfall well is challenging, but one way of doing so is through a discrete mixture of, say $Z=3$, distributions. These can be interpreted as rainfall severity states, i.e. ``dry'', ``wet'', ``wetter''. A discrete random quantity $z_t\in \{1,2,3\}$ is used to characterize the distribution of rainfall $y_t$ as:
\begin{equation}
f(y_t) = \sum_{j=1}^Z\mathds{1}(z_t=j)f(y_t|z_t)
\end{equation}
where $\mathds{1}$ is the identity function, and $p(y_t|z_t)$ is the conditional distribution of rainfall $y_t$ for each state. HMMs allow for temporal dependence by assuming that $z_t$ is an unobserved discrete Markov chain, so that temporal structure is introduced in the persistence of each state. This is parametrised by a transition matrix $P=\{p_{i,j}\}$ where $p_{i,j}=\mbox{Pr}(z_t=j|z_{t-1}=i)$.

Such models are conventionally homogeneous, meaning that the transition between states in the HMM is time invariant. However, this does not allow for the effect of seasonal variation or climatological covariates on the temporal structure of rainfall. Several articles (such as \cite{nh_bayesian}, \cite{nh_forecast}, \cite{nh_alcohol}) have instead presented non-homogeneous hidden Markov models (NHMMs), where covariates are used to characterise the parameters of the transition matrix. This added flexibility could be used to allow for seasonal or long term  heterogeneity in the temporal structure of rainfall.

Additionally, HMMs are limited by the fact that the number of time steps the hidden variable $z_t$ persists in a given state is implicitly Geometrically distributed. The right plot of Figure \ref{fig:data_quantiles} illustrates that Exeter's dry period length distribution has a very heavy tail, with several dry periods lasting hundreds of hours. This would be a concern if the HMM consists of only one dry state and hence relies on an implicit Geometric model to capture this distribution, which we illustrate by also plotting a method of moments Geometric fit to the dry period lengths. 

Including additional unique `dry states', which the hidden state parameter could transition between, may introduce sufficient flexibility to capture the longest dry periods. However, this would impede the physical interpretability of the model and potentially introduce identifiability issues. If it is the case that the dry period distribution is seasonally structured, then some improvement may be possible by introducing non-homogeneity to the dry state, though this may still be insufficient. 

A more potent solution would instead be to use a hidden semi-Markov model (HSMM), where the persistence distribution is explicitly defined and thus can be chosen to have a heavier tail. However, HSMMs are often impractical and too computationally expensive to implement. This is especially true when the total number of time steps $T$ is large, as many of the implementation methods have an algorithmic order $O(Z^2 T^2)$ or even $O(Z^2 T^3)$, compared to only order $O(Z^2 T)$ for HMMs. Often, it is necessary to restrict the upper bound of the support of the persistence time distribution prior to fitting the model, to ensure computational feasibility \citep{Economou2014}. However, this strong prior statement about the persistence distribution can lead to invalid parameter estimates \citep{Dewar2012}. While this issue can be overcome by making the restriction adaptive in the implementation process, the method is still of far greater computational complexity than the basic HMM.

Once an appropriate choice of temporal structure is made, it remains to specify the conditional rainfall model. This usually involves the mixture of a Bernoulli quantity, which represents the occurrence of rainfall, and a strictly positive quantity, to represent the intensity of rainfall, conditional on occurrence. The choice of distribution for the positive rainfall values is made more difficult by the presence of extremely high observations. For many applications of a statistical rainfall model, including urban flood modelling, the risk posed by these extreme events are of particular concern. \cite{extremecompare} argue that the commonly used Gamma distribution is not able to capture these extremes well. While some authors prefer to use the Weibull distribution \citep{sardinia}, \cite{extremecompare} also concluded that this distribution is lacking.

Several approaches, such as \cite{hybriddist} aim to better capture extremes by mixing a more typical distribution, in this case the Exponential distribution, with the Generalised Pareto distribution for values above a given threshold, which is estimated by imposing a continuity constraint on the two distributions. However, \cite{extremecompare} also found that these approaches, while still performing better than the Gamma and Weibull distributions, are not able to capture well the likelihood of extreme values. 


In what follows, we present an extended HMM-based framework that is flexible enough to adequately capture both temporal persistence and extremes in rainfall, while retaining interpretability.

\subsection{Clone states and non-homogeneity}
We begin by considering the basic three-state HMM introduced earlier. One state is intended to capture dry periods, while the remaining two are intended to capture wet periods. These wet states may end up representing periods of low and high rainfall intensity, respectively, or may differ in how long they last or how often they occur. The HMM structure for $z_t$ is defined by an initial state probability vector $P_0$ and a transition matrix $P$:
\begin{eqnarray}
P&=&\begin{pmatrix}
p & q_1 (1-p) &  q_2 (1-p) \\
r_{1,1} & r_{1,2} &  r_{1,3} \\
r_{2,1} & r_{2,2} &  r_{2,3}
\end{pmatrix}
\end{eqnarray}
For reasons that will soon be clear, we parametrise the first row (corresponding to the dry state) in terms of the probability of remaining in the dry state ($p$), and the conditional (on a transition out of the dry state) probabilities of transitioning into each wet state, $q_1$ and $q_2$ such that $q_1+q_2=1$. As discussed in Section \ref{sec:background}, a restriction of this model is that the length of time spent in the dry state has an implicit Geometric($p$) distribution, which may not be sufficiently flexible to capture long dry periods. For a long time series, such as $t=1,\ldots,70128$ hours in our application later on, an HSMM framework is prohibitively computationally expensive, so here we look for ways to retain the practicality of HMMs while making them more flexible in terms of capturing long persistence periods.

\cite{Courgette} present a way of achieving a more flexible persistence distribution for a given state, without losing the convenience the HMM framework. For clarity of exposition, denote the dry state as $d$ and the wet states as $w_j$ for $j=1,2$. The idea is to introduce a number of ``clone'' dry states $d_1,\ldots,d_D$, which are all identical to each other and to the original dry state $d$, in the sense that they all have the same conditional model. The transition matrix is then defined in such a way that transitions from $w_j$ are only possible to the first clone state $d_1$. From here, the hidden chain can persist in the first clone state $d_1$ with probability $p_1$, or transition to $w_j$, or transition to the second clone state $d_2$. If the chain transitions to $d_2$, it can remain there with probability $p_2$, or transition to $d_j$ or to the next clone state $d_3$, and so on. The motivation behind this approach is that, while the number of time steps spent in each of the clone states is still Geometric, the total amount of time spent in any of the clone states before transitioning to another unique state $w_j$ is a more flexible distribution -- essentially a weighted sum of Geometric distributions. Note that the dry state $d$ is now only implicitly defined in the sense that all clone states relate to the same conditional model for ``dryness'' (low rainfall). 

Here we also employ an approach that is based on the introduction of clone states. However, our formulation is such that it allows for modelling flexibility, particularly in terms of introducing temporal non-stationarity in the persistence of the dry state. To that end we consider the following constrained transition matrix:
\begin{eqnarray}
P=\begin{pmatrix}
p_1  & . & 0 & q_1 (1-p_1) &  q_2 (1-p_1) \\
.  & . & . & . & .  \\
0  & . & p_D & q_1 (1-p_D) &  q_2 (1-p_D) \\
v_1  r_{1,1}&  . & v_D  r_{1,1} & r_{1,2} &  r_{1,3} \\
v_1  r_{2,1}&  . & v_D  r_{2,1} & r_{2,2} &  r_{2,3}
\end{pmatrix}
\end{eqnarray}
Here, transitions are possible from wet states $w_j$ into any of the clone dry states $d_i$, while no transitions between the clone states are possible. The latter is achieved by constraining the off-diagonal entries of the first $D$ rows and columns to be zero. The result of this is that, while the dry state persistence distributions are each Geometric($p$), $p$ can now be thought of as a (categorical) random quantity taking values in $\{p_1,p_2,p_3\}$, such that the marginal distribution for the time spent in the implicit dry state is, like the approach in \cite{Courgette}, a more flexible Geometric mixture.

To ensure that it is possible to conceptualise the clone dry states as a single state, further constraints are imposed on the transition matrix: First, conditional on a transition from a dry state to a wet state, the transition probabilities ($q_1$ and $q_2$) into each wet state are invariant of the dry state. Second, conditional on a transition from a wet state to a dry state, the transition probabilities ($p_1,...,p_D$) into each dry state are invariant of the wet state. These are in addition to the constraint that the conditional model for rainfall occurrence and intensity is the same for all of the dry states.

This approach is equivalently flexible to the one from \cite{Courgette}, in the sense that it can better capture heavy tailed persistence distributions. This implies that, without sacrificing the physical interpretability of having only one dry state, or the practicality of the HMM framework, extra flexibility is afforded to potentially capture better the longest dry periods. However, as the parameters of the transition matrix are time-constant, the model can't capture seasonal or annual variation in the expected length of dry periods which may be, for example, longer on average in the summer than in the winter. The advantage of our approach is that it is straightforward to directly model the dry state persistence probabilities $p_1,...,p_D$ as temporally-varying.
One way of achieving this, as defined in \eqref{eq:logit_persistence}, is a logistic model for the dry state persistence probabilities. Here $u(t,d)$ represents a general model of time $t= 1,...,T$ and hidden state $d=1,...,D$, which may also include covariate effects, including large-scale climate indices such as the North Atlantic Oscillation (NAO).
\begin{equation}
\log \left(\frac{p_d(t)}{1-p_d(t)}\right) = u(t,d) \label{eq:logit_persistence}
\end{equation}
For our application to the Exeter data, we characterise u(t,d) by combining an intercept term $\iota(d)$ which is different for each clone dry state, and two penalised splines which are common across the states:
\begin{equation}
u(t,d) = \iota(d) + a_1(t)+a_2(t) \label{eq:persistence}.
\end{equation} 
The first spline, $a_1(t)$, is a cyclic (the two end points have equal value) cubic spline of the time of year, which is intended to capture seasonal variation in the length of dry periods. The second, $a_2(t)$, is a cubic spline of time overall, which is intended to capture between-year variation in the average length of dry periods. The use of splines affords the model flexibility to capture different dry period persistence structures which may occur in different climatic conditions, and also to better capture very long dry periods.

\subsection{Conditional rainfall model}
Having specified a non-stationary (non-homogeneous) and essentially semi-Markovian latent structure, it remains to define a conditional model for rainfall occurrence and intensity. First recall the notation $z(t)$, the latent state at any given time point. Continuing with our three state example, $z(t)$ takes only 3 values (dry, wet, wetter), noting that the dry state is made up of $D$ clone dry states, among which the conditional model is the same.

As zero rainfall is generally a common  observation (approximately 88\% of all observations in the Exeter time series), it makes sense to mix a continuous distribution for rainfall intensity with a probability mass at zero. This probability of zero rainfall, $\pi_t$, should vary with the latent state $z_t$, for example it should be higher in the dry state than in the wet states, and it may also vary with time and/or depend on climatological covariates. We achieve this by employing a logistic model for $\pi_t$:
\begin{equation}
\log\left(\frac{\pi_t}{1-\pi_t}\right) = v(t,z_t)  \label{eq:logit_zero}
\end{equation}
For our application to the Exeter gauge, we once again employ a combination of a seasonal and an overall temporal spline:
\begin{equation}
v(t,z_t) = \eta(z_t) + b_1(t,z_t)+b_2(t,z_t)
\end{equation} 
Unlike the splines for the dry state persistence probabilities, splines $b_1(t,z_t)$ and $b_2(t,z_t)$ are different and independent across each of the three states.

As discussed in Section \ref{sec:conventional}, there are many choices for the distribution of rainfall intensity, including Gamma, Weibull, Log-Normal and hybrid distributions. One of the key advantages of the approach we advocate is that it is possible to choose from any of these or other distributions, even using different distributions for each state if desired. Recalling that one of our key modelling aims is to capture extreme values well, we opt for the zero-location (zero-threshold) Generalized Pareto distribution (GPD), with scale $\sigma_t$ and shape $\xi_t$:
\begin{eqnarray}
\log(\sigma_t) &=& \alpha(z_t) + c_1(t,z_t)+c_2(t,z_t), \\
\xi_t &=& \gamma(z_t) + d_1(t,z_t)+d_2(t,z_t).
\end{eqnarray}
Once more we make use of seasonal ($c_1(t,z_t)$ and $d_1(t,z_t)$) and temporal ($c_2(t,z_t)$ and $d_2(t,z_t)$) splines to capture inhomogeneity. By including independent splines for each state in all of the parameters of the conditional model (in this case $\pi_t$, $\sigma_t$ and $\xi_t$), a high degree of flexibility for capturing seasonal and temporal variation in the rainfall distribution is afforded.

In our dataset, the hourly observations are rounded to the nearest 0.2mm, which means that the likelihood should be adjusted accordingly. For example, if a rainfall observation is 2mm, the contribution to the likelihood should not just be the density $f(2mm;...)$, but should instead be $P(1.9mm < X \leq 2.1mm)$. Furthermore, we truncate the GPD at 0.1mm such that the zero probability accounts for all values less than 0.1mm (which would be rounded to zero), such that the complete density function is:
\begin{eqnarray}
f(x;\pi,\sigma,\xi)&=&\begin{cases} 
      \pi & x=0 \\
      (1-\pi)\frac{F(x+0.1)-F(x-0.1)}{1-F(0.1)} & x=0.2,0.4,... 
   \end{cases} \label{likelihood} \\
   F(x;\sigma,\xi) &=& 1-\left(1+\frac{\xi x}{\sigma} \right) ^{-\frac{1}{\xi}}
\end{eqnarray}   
where $F$ is the cumulative distribution function of the zero-location GPD.

\subsection{Prior distributions and implementation}
We apply the model to hourly observations from the Exeter International Airport rainfall gauge, a time series of 70128 values spanning the 8-year period 2010 to 2017. 

To keep the model as general as possible, we specified uniform Dirichlet($\bm{1}$) prior distributions for transition matrix parameters $\bm{v}$, $\bm{r}_1$ and $\bm{r}_2$. We also specified non-informative Normal($0,10^2$) prior distributions for the intercept parameters $\iota(d)$, $\eta(z_t)$, $\alpha(z_t)$ and $\gamma(z_t)$, where we use $d=1,2,3$ clone dry states. However, a common problem with hidden Markov models is label switching, where the conditional models of one or more states swap. When this happens, the overall model is the same but parameter inference is convoluted, especially in a Bayesian implementation where Markov Chain Monte Carlo (MCMC) is employed. To prevent this, we impose the following constraints on the intercept parameters:
\begin{eqnarray}
\iota(1) &>& \iota(2) > \iota(3)\label{eq:iota_constraint}\\
\eta(1) &>& max(\eta(2),\eta(3))\label{eq:eta_constraint}\\
\gamma(3) &>& \gamma(2)\label{eq:gamma_constraint}
\end{eqnarray}
Constraints \eqref{eq:iota_constraint} and \eqref{eq:eta_constraint} do not really restrict the model, they just order the clone dry states and specify that the dry states should have a higher average probability of zero rainfall than the two wet states, respectively. The third constraint \eqref{eq:gamma_constraint}, that the second wet state should have a heavier tail behaviour on average, regardless of the scale parameter, can be viewed as a potential restriction, but can be justified subject to satisfactory model checking performance.

All splines were set up using the \texttt{jagam} function in the \texttt{mgcv} package for the programming language R \citep{R}. We specified 6 equidistant knots for the seasonal splines and 1 knot for each year (8 in total) for the overall temporal splines. For each spline, the coefficients are assigned Multivariate-Normal priors \citep{jagam} where the covariance matrix is scaled by parameter $\nu$ (unique for each spline), which acts as a smoothing penalty (where smaller values of $\nu$ correspond to a stricter penalty). More generally, this penalty is intended to avoid over-fitting, but here we would like our spline effects to be quite smooth as they are intended to capture long-term variation, while the HMM latent state $z_t$ is intended to capture short to medium term variation. For these penalty parameters ($\nu$), we specified Half-Normal($0,\sqrt{2}^2$) prior distributions, which corresponds to a modest smoothness penalty.

The model was implemented using the \texttt{nimble} package \citep{nimble}, a comprehensive suite for flexible  MCMC inference. In this case, we needed to create a custom likelihood function for the rainfall observations which incorporates a version of the recursive forward algorithm used to compute the marginal likelihood in HMMs \citep{Scott:2002} and thus avoiding sampling the latent states. This was adapted to allow for a temporally-varying transition matrix. We ran four MCMC chains for a total of 20k iterations, discarding the first 10k as burn-in. Owing to the complexity and size of the model (70128 hours of data), the model takes 2-3 days (on an Intel Core i9-7900X processor with 64GB of memory) with the four chains running in parallel. Each chain was randomly initialised at different parameter values and was assigned a different random number generator seed. Convergence of the four chains was assessed by visual inspection of trace plots and by computing the Potential Scale Reduction Factor (MPSRF) \citep{convergence} for all of the following parameters: the initial state probability $P_0$; static transition matrix parameters ($\bm{q}$, $\bm{v}$ and $\bm{r}$), all of the intercepts ($\bm{\iota}$, $\bm{\eta}$, $\bm{\alpha}$ and $\bm{\gamma}$); and all of the spline coefficients. This metric compares the variance between the chains to the variance within the chains. If the two variances are similar for a given parameter then this typically results in a PSRF close to 1. Starting from different initial values and obtaining a PSRF close to 1 (less than 1.05 by convention) gives the best indication that the chains have converged to the posterior. Here, the median PSRF among this set of parameters was 1.01, with a mean of 1.02, suggesting the chains have converged. All of the code to prepare and run the model is provided as supplementary material.

\section{Model Checking and Results}\label{sec:check}
In this section we illustrate the model's performance through comprehensive model checking and analysing some key results. We do the former by simulating a new time series of length 70128, $\tilde{\bm{r}}$, from each posterior sample. Doing this, we have 4000 simulated time series (after thinning the posterior samples by 10) and the general principle is then to assess whether certain characteristics of the observed values $\bm{r}$ are extreme relative to simulations from the model \citep{Gelman2013}.

\subsection{Temporal structure}
\begin{figure}[ht!]
\floatbox[{\capbeside\thisfloatsetup{capbesideposition={right,center},capbesidewidth=0.5 \linewidth}}]{figure}[1\FBwidth]
{\caption{Median simulated 800 longest sorted dry periods, defined as periods where rainfall does not exceed 0.2mm in any one hour, compared to the 800 longest observed periods and with associated 95\% prediction intervals.}\label{fig:dry_persistence}}
{\includegraphics[width=\linewidth]{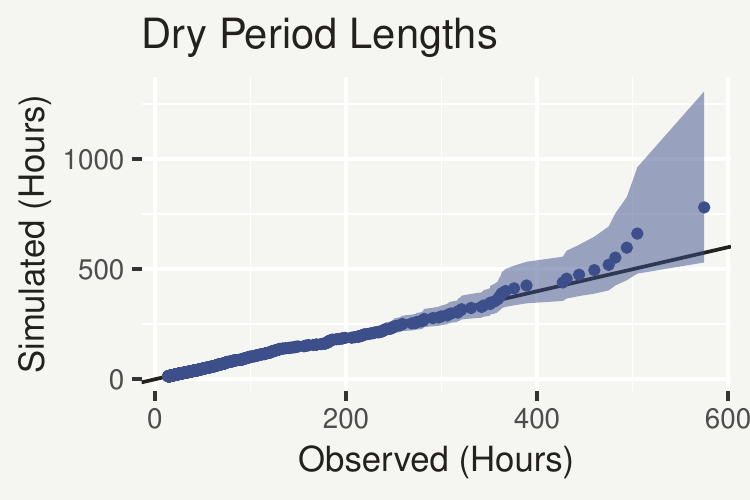}}
\end{figure}

We begin by assessing the model's ability to capture long dry periods, one of the three key characteristics of rainfall data we identified as important in Section \ref{sec:intro}. We do this by, for each simulated time series $\tilde{\bm{r}}$, calculating the length of all dry periods (defined as periods where rainfall does not exceed 0.2mm in any hour), sorting them into ascending order and storing the 800 greatest lengths (as there is a different number of dry periods in each simulated time series). Figure \ref{fig:dry_persistence} shows the median predicted 800 greatest dry period lengths, with 95\% prediction intervals, compared to the 800 greatest dry period lengths in the observed data. The model generally does an excellent job of capturing the distribution of dry period lengths, with the median values tracking the observed values (diagonal line) very closely, all the way up until the last few points, which are still contained within the 95\% prediction intervals. 

In developing our approach, we found that the inclusion of `clone' dry states made a dramatic improvement over the baseline HMM in capturing this distribution well (the baseline was not able to capture any part of this distribution remotely well), but it wasn't quite good enough until we allowed the dry persistence probabilities to vary with time.
\begin{figure}[ht!]
\begin{center}
\includegraphics[width=\linewidth]{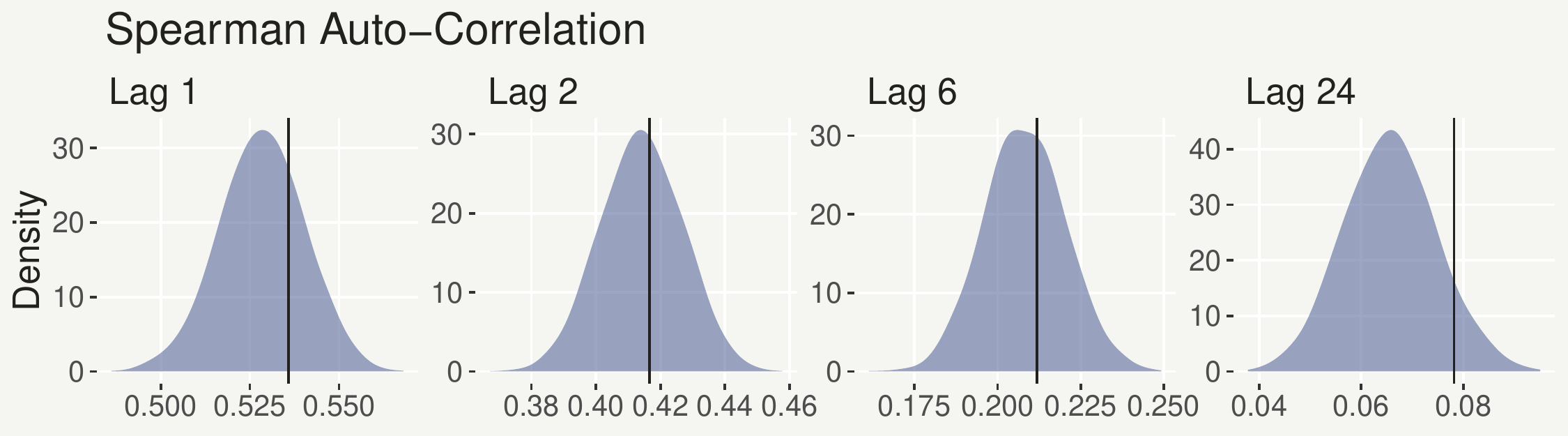}
\end{center}
\caption{Density plots of simulated Spearman autocorrelations from the model, with vertical lines representing the corresponding observed autocorrelations.}\label{fig:autocor}
\end{figure}

We move on to assessing the temporal structure more generally, which we do by calculating the Spearman auto-correlation, at various lags, for each simulated time series $\tilde{\bm{r}}$. Here, the lag-1 autocorrelation means the correlation between rainfall values one hour apart (the correlation between the two vectors $\tilde{r}_1,...,\tilde{r}_{70127}$ and $\tilde{r}_2,...,\tilde{r}_{70128}$), lag-2 is the correlation between values two hours apart and so on. We can then compare the corresponding autocorrelations for the observed data to the distribution of simulated autocorrelations. Figure \ref{fig:autocor} shows density estimates of simulated lag-1, lag-2, lag-6 and lag-24 autocorrelations. None of the observed autocorrelations are extreme with respect to the distributions of simulated statistics, suggesting that the model is generally capturing the temporal structure well.

\subsection{Seasonal distributions}
Whilst capturing the whole distribution of hourly rainfall values well, including extremes, can be challenging in itself, an even greater challenge is capturing this distribution as it varies by season. For example, many models overestimate extremes in the winter and underestimate them in the summer \citep{rainfall_uncertainty}. 
\begin{figure}[ht!]
\begin{center}
\includegraphics[width=\linewidth]{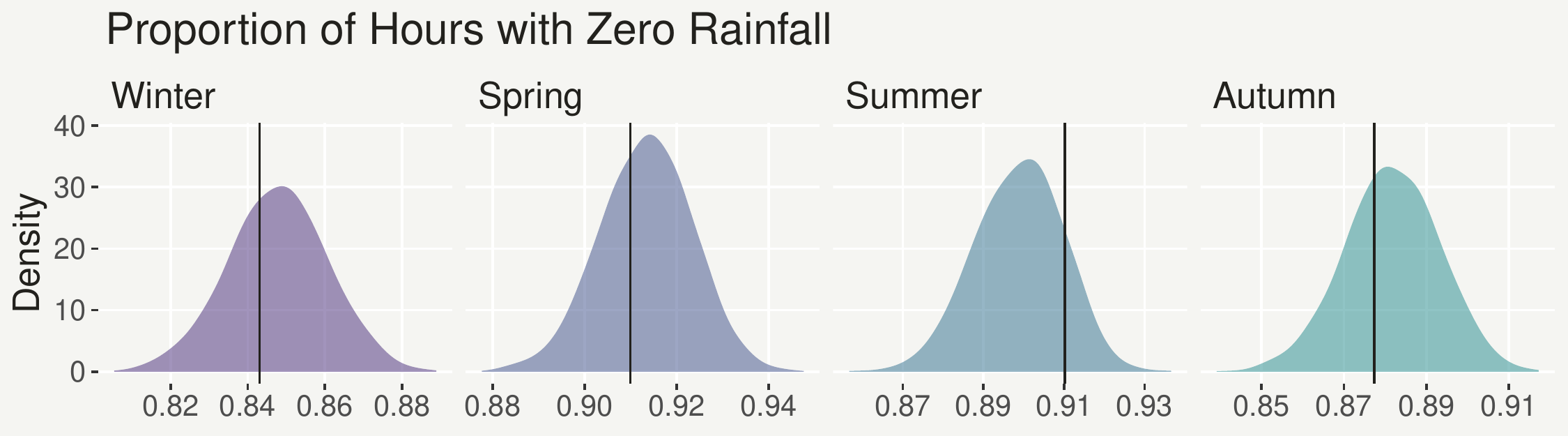}
\end{center}
\caption{Median predicted sorted rainfall values, from 4000 simulated 8-year time series, with associated 95\% credible intervals.}\label{fig:season_zero}
\end{figure}

First we check that the model is able to capture seasonal variation in the occurrence in rainfall. Figure \ref{fig:season_zero} shows density plots of the proportion of zero values in each calendar season from the simulated time series $\tilde{\bm{r}}$, compared to the proportions in the observed values. All of the observed values are captured quite well, for example the model captures the increased proportion of zeros in the summer compared to the winter, so it's clear that the model is able to reproduce seasonal variation in rainfall occurrence.
\begin{figure}[ht!]
\begin{center}
\includegraphics[width=\linewidth]{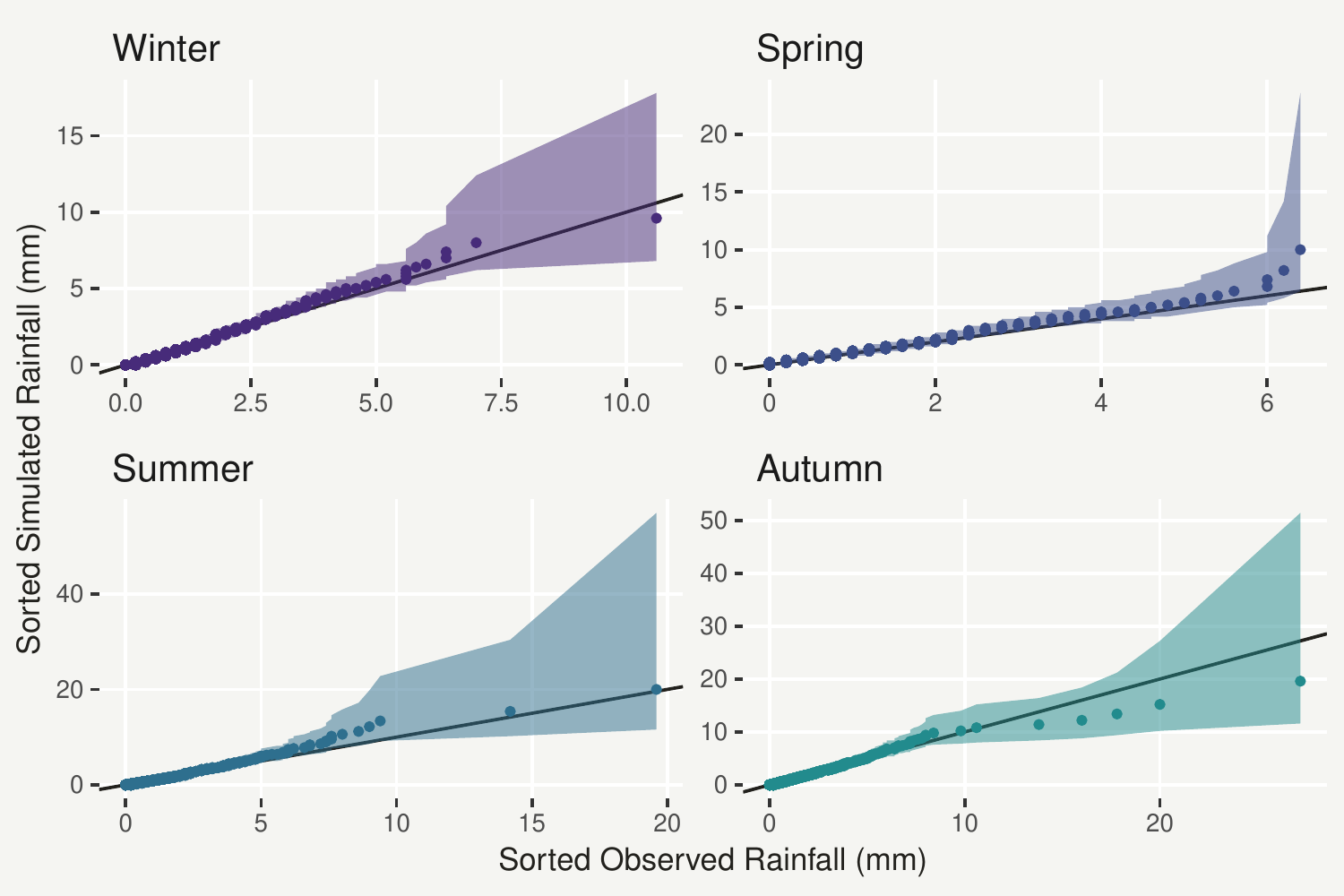}
\end{center}
\caption{Median predicted sorted rainfall values, from 4000 simulated 8-year time series, by calendar season and with associated 95\% prediction intervals.}\label{fig:season_sorted}
\end{figure}

Next we check whether the model is able to capture seasonal variation in rainfall intensity. We do this for each season, by sorting (ranking) each simulated rainfall time series $\tilde{\bm{r}}$, and then compare the distribution of intensities for each rank to the observed ranked intensity (this has a similar interpretation to a quantile-quantile plot). Figure \ref{fig:season_sorted} compares the observed sorted values to the median simulated sorted values, with associated 95\% prediction intervals. Looking at the plots, it's clear that the distribution of rainfall values varies greatly by season, with higher (and more extreme) rainfall values in the summer and the autumn than in the winter and the spring. Despite this variation, the model is able to capture each season's distribution very well, all the way up to the extremes, especially given only two wet states were used. Equivalent return level plots can be found in the supplementary material.

\subsection{Daily and monthly rainfall checking}\label{app:daily_monthly}
For some applications it may also be important that the model captures well the cumulative rainfall over a longer time period, such as a day, or a month. To assess this, we aggregate both the observed time series $\bm{r}$ and each simulated time series $\tilde{\bm{r}}$ (as detailed in Section \ref{sec:check}) into daily values and monthly values. We can now assess whether the aggregated data ($\bm{r}_{daily}$ and $\bm{r}_{monthly}$) is extreme with respect to the simulated data ($\tilde{\bm{r}}_{daily}$ and $\tilde{\bm{r}}_{monthly}$).
\begin{figure}[ht!]
\begin{center}
\includegraphics[width=\linewidth]{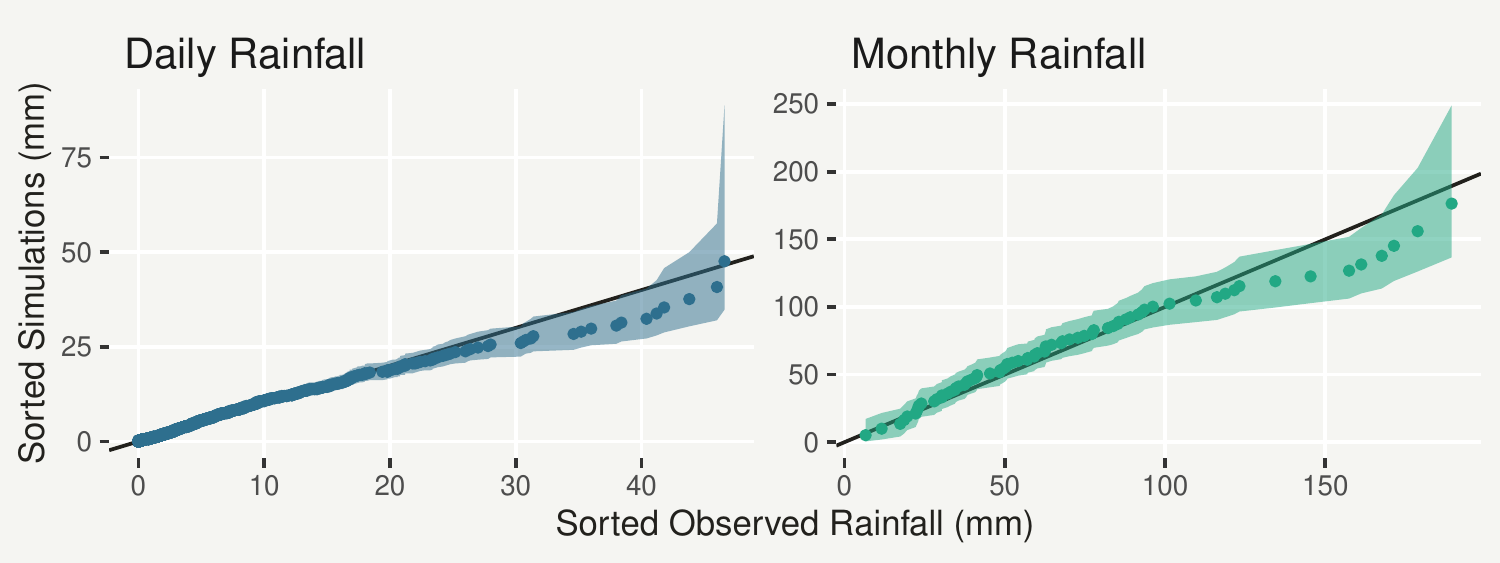}
\end{center}
\caption{Median sorted simulated daily (left) and monthly (right) rainfall values, with 95\% prediction intervals, compared to the corresponding sorted observed values.}\label{fig:daily_monthly_sorted}
\end{figure}

First we check that the model captures the distribution of daily rainfall intensity well. The left panel of Figure \ref{fig:daily_monthly_sorted} shows predicted sorted values from the model, compared to the sorted observed values. The model captures most of the distribution excellently, though there is some deviation in the very upper tail. In this region, the observed values are either contained by or extremely close to the 95\% prediction intervals, and the highest values are actually captured very well, so we believe the model is doing well enough. 

Similarly, we can also check the model captures the distribution of monthly rainfall intensity well. Once again, as illustrated in the right panel of Figure \ref{fig:daily_monthly_sorted}, the model captures most of the distribution virtually perfectly, with some mild deviation in the very upper tail but not to a concerning extent.

\subsection{Temporal effects}
\begin{figure}[ht!]
\begin{center}
\includegraphics[width=\linewidth]{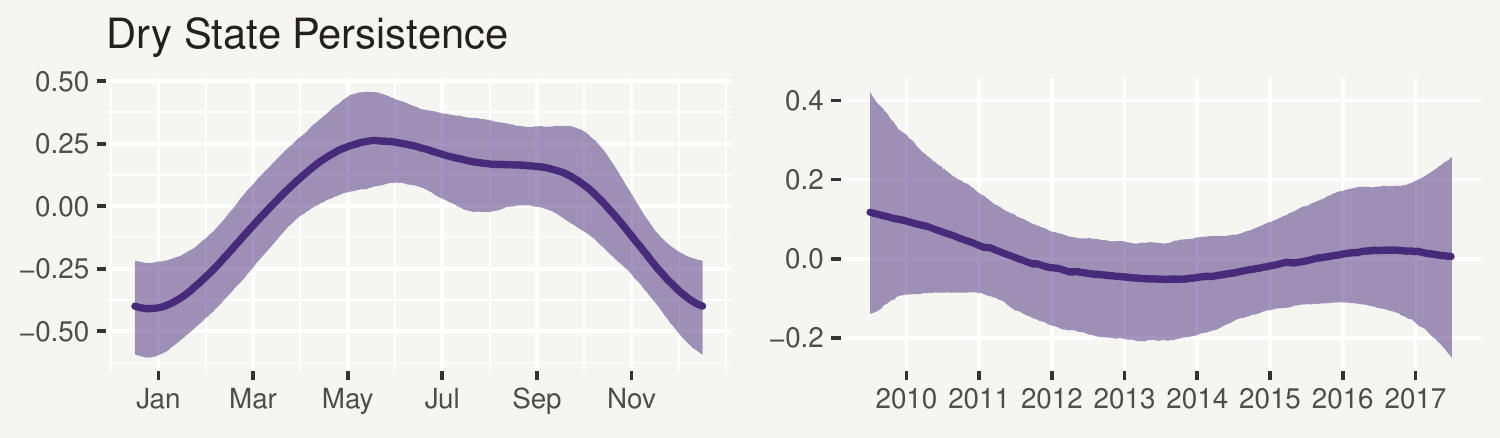}
\end{center}
\caption{Posterior median predicted effect of the time of year (left) and overall time (right) on the dry state persistence probabilities, with 95\% credible intervals.}\label{fig:persistence_spline}
\end{figure}

As well as aiding in fitting the data better, the temporal effects can indicate how different characteristics of rainfall vary with time of year and between years. Figure \ref{fig:persistence_spline} shows the posterior median predicted effects of time of year (left) and time overall (right) on the expected persistence of dry periods, as defined by \eqref{eq:persistence}. Looking at the cyclic effect of time of year, there is overwhelming evidence that dry periods are longer in the warmer months (May-September) than the rest of the year. Looking at the overall temporal effect, there isn't strong evidence of any change in the expected length of dry periods over the 8 years.

\begin{figure}[ht!]
\begin{center}
\includegraphics[width=\linewidth]{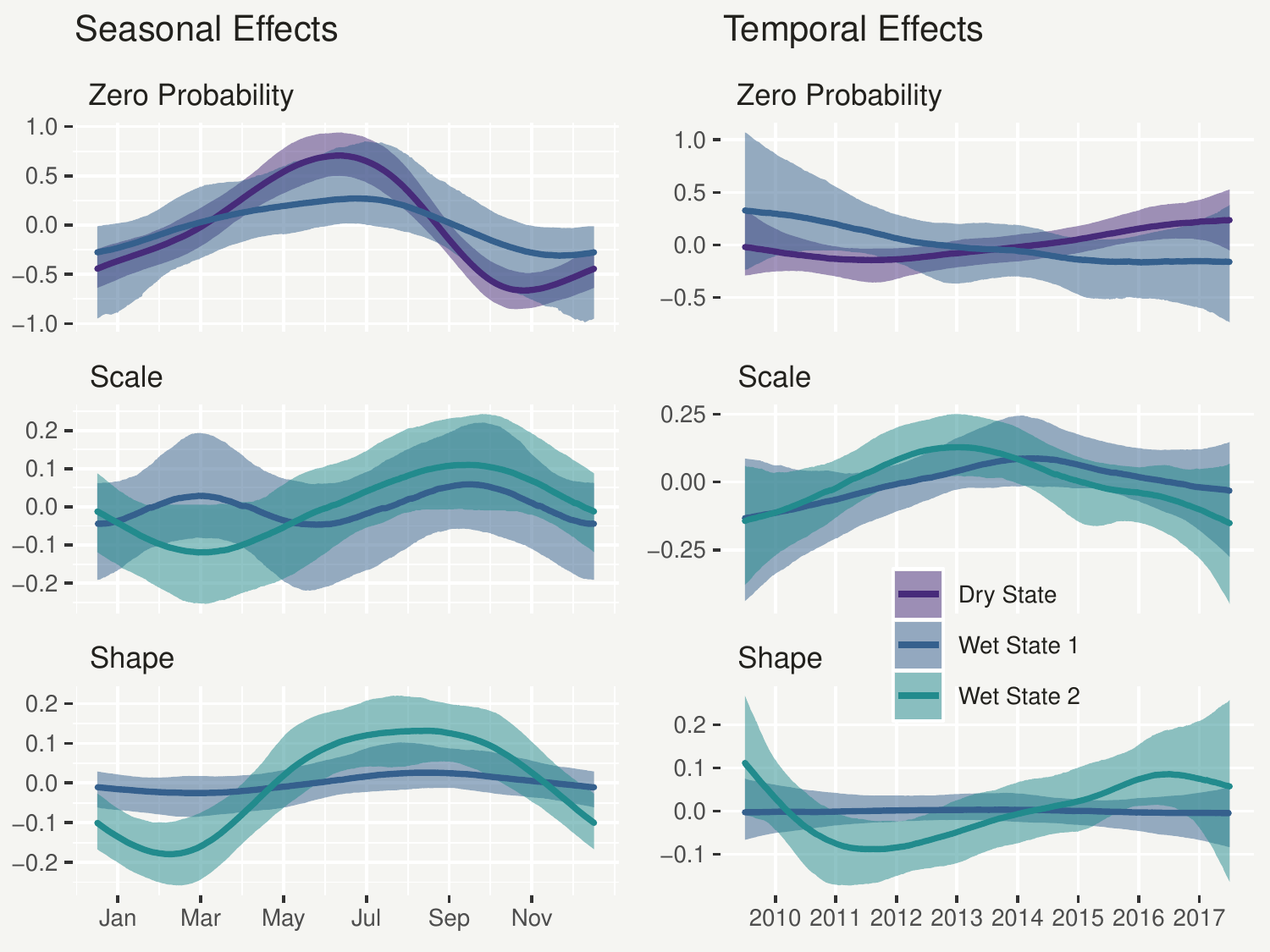}
\end{center}
\caption{Posterior median predicted effects of time of year (left column) and overall time (right column) on the conditional probability of zero rainfall (top row), and the Generalized Pareto scale and shape parameters (central and bottom rows, respectively), by hidden Markov state and with 95\% credible intervals.}\label{fig:conditional_splines}
\end{figure}

Figure \ref{fig:conditional_splines} shows the effects of time of year (left column) and overall time (right column) on the conditional probability of zero rainfall (top row) and the distribution of rainfall intensity, through the Generalized Pareto scale and shape parameters (central and bottom rows, respectively). First note that the zero probability spline for wet state 2 isn't plotted, because the zero probability in wet state 2 was so low that the spline had no predicted effect and was very uncertain. Similarly, the dry state converged to a situation where it essentially produced exclusively values of 0mm and 0.2mm, meaning the splines for the scale and shape parameters also had no predicted effect.

Looking at the cyclic effects, there is a clear pattern where the conditional probability of zero rainfall is higher in the warmer months in both the dry state and wet state 1. This is in addition to the longer dry periods seen at this time of year in Figure \ref{fig:persistence_spline}. Now looking at the effects for wet state 2, we can see that both the shape and scale are higher in the summer and autumn than elsewhere in the year, meaning both higher average rainfall values and a heavier tail behaviour. Interestingly, the scale parameter is at it's highest in the autumn, while the shape parameter is highest in the summer. This makes sense to us because our anecdotal experience of Exeter is that autumn is more consistently rainy, while summer is prone to intermittent downpours. The seasonal intensity  effects for wet state 1 don't show a similarly strong pattern, indicated by the 95\% credible intervals containing zero.

The overall temporal effects show several strong (in terms of 95\% credible interval certain) trends over the 8-year period. For example, the scale parameter of wet state 2 is highest around 2013, which may correspond to the occurrence of severe storm events around this time (e.g. \cite{EA1314}). These effects seem to be capturing real trends, rather than just sampling variation between the years, though a longer time series may be necessary to investigate whether they relate to large scale climate indices like the NAO.

\section{Discussion}\label{sec:discuss}
In this article we discussed the role of stochastic or statistical rainfall modelling in the context of hydrological applications, such as urban flood modelling. We illustrated how the flexibility of the hidden Markov model framework allowed us to construct a comprehensive model for sub-daily rainfall, which is able to capture all of the following crucial features: seasonal variation in rainfall occurrence and intensity, long dry periods and extreme values. Our model incorporates several innovations compared to conventional approaches. These include clone states and temporal non-homogeneity in the transition matrix, which together allow the model to capture even the longest dry periods. Set in the Bayesian framework, our model also allows for a rich quantification of parametric and predictive uncertainty, meaning we can use posterior predictive checking to verify our model captures important characteristics of the data. In addition, the application to hourly rainfall data illustrates the applicability of the model in situations with high (temporal) resolution, something that is noticeably absent from the literature on direct rainfall models. 

To demonstrate the effectiveness of our approach, we applied a relatively simple model comprising 3 clone dry states and 2 wet states to an 8-year long time series of hourly values from a rainfall gauge in Exeter, UK. We found that the model is able to capture well the distribution of dry period lengths, seasonal variation in occurrence and intensity (including extreme values) and the distribution of intensity when aggregated to daily and monthly resolutions. We also illustrated how the model output can be interpreted in terms of how the rainfall occurrence and intensity change over the course of the year and over the whole time period.

We opted to apply the model to the particular time series from Exeter as it is situated in a region where floods pose a real risk to society, and because of the presence of several extreme values (arising from severe storms) which make modelling challenging. The inclusion of splines in every part of the model affords a high degree of flexibility, in the sense that the states can change completely for different times of year and between years. That said, it is possible that there are some climates where the specific model we used to illustrate the framework may not be sufficiently flexible. In this case, the advantage of our approach is that it is fairly trivial to add more wet states, use alternative conditional distributions, combine different conditional distributions, to name just a few potential adaptations. However, as with many statistical endeavours, this comes at the cost of increased complexity, so a balance must be struck to find a model which performs well enough without being impractical.

Finally, in this article we have focussed on modelling the time series from one spatial location. To cater for applications where simulations at more than one location are required, future research will involve combining the innovations presented in this article with methods such as coupled hidden Markov models \citep{coupled}, so that dependence between multiple spatial locations can be allowed for.

\section*{Supplementary Material}
All R code required to implement the model is provided as supplementary material. Hourly rainfall data from the Exeter International Airport gauge is available upon registration at \url{http://dx.doi.org/10.5285/7aaa582fb00246b794dc85950f1be265}.

\section*{Acknowledgements}

This work was funded in part by a Natural Environment Research Council GW4+ Doctoral Training Partnership studentship [NE/L002434/1].

\bibliography{rainfall}

\end{document}